\documentclass[graybox,natbib,nosecnum]{svmult}
\bibpunct{(}{)}{;}{a}{}{,} 

\pdfoutput=1   
\usepackage{ amssymb }
\usepackage{mathptmx}       
\usepackage{helvet}         
\usepackage{courier}        
\usepackage{type1cm}        

\usepackage{makeidx}         
\usepackage{graphicx}        
\usepackage{multicol}        
\usepackage[bottom]{footmisc}
\usepackage[normalem]{ulem}	
\usepackage{hyperref}  

\usepackage{soul}   
\usepackage{xcolor}

\makeindex             

\begin{document}

\title*{Signatures of Star-Planet Interactions}
\author{Evgenya L. Shkolnik and Joe Llama}
\institute{Evgenya L. Shkolnik \at ASU School of Earth and Space Exploration, Tempe, AZ 85287. USA, \email{shkolnik@asu.edu}
\and Joe Llama \at Lowell Observatory, 1400 W. Mars Hill Road, Flagstaff, AZ. 86001. USA \email{joe.llama@lowell.edu}}
%
%
\maketitle

\abstract{Planets interact with their host stars through gravity, radiation and magnetic fields, and for those giant planets that orbit their stars within $\sim$10 stellar radii ($\sim$0.1 AU for a sun-like star), star-planet interactions (SPI) are observable with a wide variety of photometric, spectroscopic and spectropolarimetric studies. At such close distances, the planet orbits within the sub-alfv\'enic radius of the star in which the transfer of energy and angular momentum between the two bodies is particularly efficient. The magnetic interactions appear as enhanced stellar activity modulated by the planet as it orbits the star rather than only by stellar rotation. Such observations allowed for the determination of the magnetic field strengths on the surfaces of four hot Jupiters. These  vary between 20 G and 120 G, in line with scaling laws that connect the strength of the magnetic field to the internal heat flow in giant planets.   These field strengths are informative for the study of the internal dynamics and atmospheric evolution of exoplanets. The nature of magnetic SPI is modeled to be strongly affected by both the stellar and planetary magnetic fields, possibly influencing the magnetic activity of both, as well as affecting the particle environment, the migration of the planet, and even the rotational evolution of the star.  As phase-resolved observational techniques are applied to a large statistical sample of hot Jupiter systems, extensions to other tightly orbiting stellar systems, such as smaller planets close to M dwarfs become possible. In these systems, star-planet separations of tens of stellar radii begin to coincide with the radiative habitable zone where planetary magnetic fields are likely a necessary condition for surface habitability.\\}

\noindent \textit{Chapter in the revised 2024 edition of the Handbook on Exoplanets by Springer Reference.}

\section{Introduction }
Giant planets located $<0.1$ AU from their parent star, also known as hot Jupiters (HJ), provide a laboratory to study the tidal and magnetic interactions between the planet and its star that does not exist in our own solar system.  These  interactions can be observed because they scale as $a^{-3}$ and $a^{-2}$, respectively, where $a$ is the separation between the two bodies.  Although HJs are rare around M dwarfs, statistics from the \textit{Kepler} survey have revealed that M stars host on average 0.24 Earth-sized planets in the habitable zone \citep{dres15}.
 We can apply the techniques trained on HJs around FGK stars on to small planet orbiting M dwarfs.

\cite{cuntz2000} first suggested that close-in planets may increase and modulate their host star's activity levels through tidal and magnetic interactions as such effects are readily observed in the comparable cases of tightly orbiting RS CVn binary systems (e.g.,~\citealt{piskunov1996,shkolnik2005b}). Variable excess stellar activity with the period of the planet's orbit rather than with the star's rotation period, indicates a magnetic interaction with the planet, while a period of half the orbit's indicates a tidal interaction.  This suggestion has spurred the search for such interactions as a means of studying the angular momentum evolution of HJ systems and as detecting the magnetic fields of exoplanets (e.g.,~\citealt{cuntz2000,shkolnik2003,saar2004,shkolnik2005,shkolnik2008,lanza2009,cohen2009}).

Exoplanetary magnetic fields provide a probe into a planet's internal dynamics and constraints on its atmospheric mass loss. This fundamental physical property of exoplanets would most directly be detected through the radio emission produced by electron cyclotron maser instability (see review by \citealt{treumann2006} and Chapter 9.6 of this book). Such emission has been detected from all of the solar system's gas giants and the Earth resulting from an interaction between the planetary magnetosphere and the solar wind. There are  no detections to date of radio emission from exoplanets although searches have been typically less sensitive at  higher emission frequencies than predicted for exoplanets (e.g.,~\citealt{farrell1999,bastian2000,lanza2009,lazio2009,jardine2008,vidotto2012} and see review by \citealt{lazio2016}).

Even though a radio detection of a planet's magnetic field ($B_p$) remains elusive, there have been confirmed detections through magnetic star-planet interactions (SPI). Nearly twenty studies of HJ systems, varying in wavelengths and observing strategy, have independently come to the conclusion that a giant exoplanet in a short-period orbit can induce activity on the photosphere and upper atmosphere of its parent star. This makes the host star's magnetic activity a probe of the planet's magnetic field.

Due to their proximity to their parent star, magnetic SPI in HJ systems can be detected because these exoplanets typically lie within the Alfv\'en radius of their parent star ($\lesssim 10R_\star$ or $\lesssim 0.1$ AU for a sun-like star). At these small separations, the Alfv\'en speed is larger than the stellar wind speed, allowing for direct magnetic interaction with the stellar surface. If the giant planet is magnetized, then the magnetosphere of the planet may interact with the stellar corona throughout its orbit, potentially through magnetic reconnection, the propagation of Alfv\'en waves within the stellar wind, and the generation of electron beams that may strike the base of the stellar corona.

In the case of characterizing habitable zone planets, the current favored targets are low-mass stars where the habitable zone is located much closer to the parent star compared to the Earth-Sun separation, making the planet easier to detect and study. Low-mass stars are typically much more magnetically active than solar-type stars. It is therefore vital that we understand how this increase in magnetic activity impacts the potential habitability of a planet orbiting close to a low-mass star and what defenses the planet has against it. In order to sustain its atmosphere, a planet around a low-mass star must be able to withstand enhanced atmospheric erosion from extreme stellar wind and also from the impact of coronal mass ejections.  Both of these reasons necessitate the push towards the detection and characterization of magnetic SPI in M dwarf planetary systems.

The need to understand magnetic SPI is also driving the modeling effort forward. There have been considerable efforts towards modeling the space weather environments surrounding close-in giant exoplanets and star-planet interactions.  
The magnetized stellar winds may interact with the close-in exoplanet  through the stars' outflows and magnetospheres, and potentially lead to observable SPI. Observing the stellar winds of stars other than the Sun is  difficult and there are very few observational constraints on the winds of low-mass stars (e.g., \citealt{wood2005}). 

Star-planet interactions can be simulated by using hydrodynamical (HD) or magnetohydrodynamical (MHD) numerical models. The modeling efforts have not only focused on studying individual systems, but have also been extended to more general scenarios to help aid the interpretation of statistical studies. MHD models for star-planet interactions require a dynamic model for the stellar corona and wind, and also a model for the planet, which acts as an additional, time-dependent boundary condition in the simulation (e.g., \citealt{cohen2011, matsakos2015,vidotto2023,strugarek2023}). The standard approach to modeling SPI involves adapting 3D MHD models originally developed for the solar corona and wind. The BATS-R-US (\citealt{powell1999,toth2012}) global MHD model forms part of the Space Weather Modeling Framework (\citealt{toth2005}) and is capable of accurately reproducing the large-scale density and temperature structure of the solar corona and has been adapted to model the winds of other stars. This MHD model uses a stellar magnetogram (or solar synoptic map) as input along with other properties of the host star, including the stellar coronal base density ($\rho$), surface temperature ($T$), mass ($M_\star$), radius ($R_\star$) and rotation period ($P_\star$). The model then self-consistently solves the ideal MHD equations for the stellar corona and wind, which in turn allows the conditions experienced by an exoplanet to be studied (e.g., \citealt{cohen2009,cohen2011,cohen2014,vidotto2009,vidotto2013,vidotto2014,doNascimento,vidotto2023, kavanagh2023}).

In this chapter, we discuss the observational evidence of magnetic SPI in FGK and M stars plus the array of models produced to explain and characterize this diagnostic, albeit complex, physical phenomenon.

\section{Planet induced and orbit phased stellar emission}

\begin{figure}
\centering
\includegraphics[width=\textwidth]{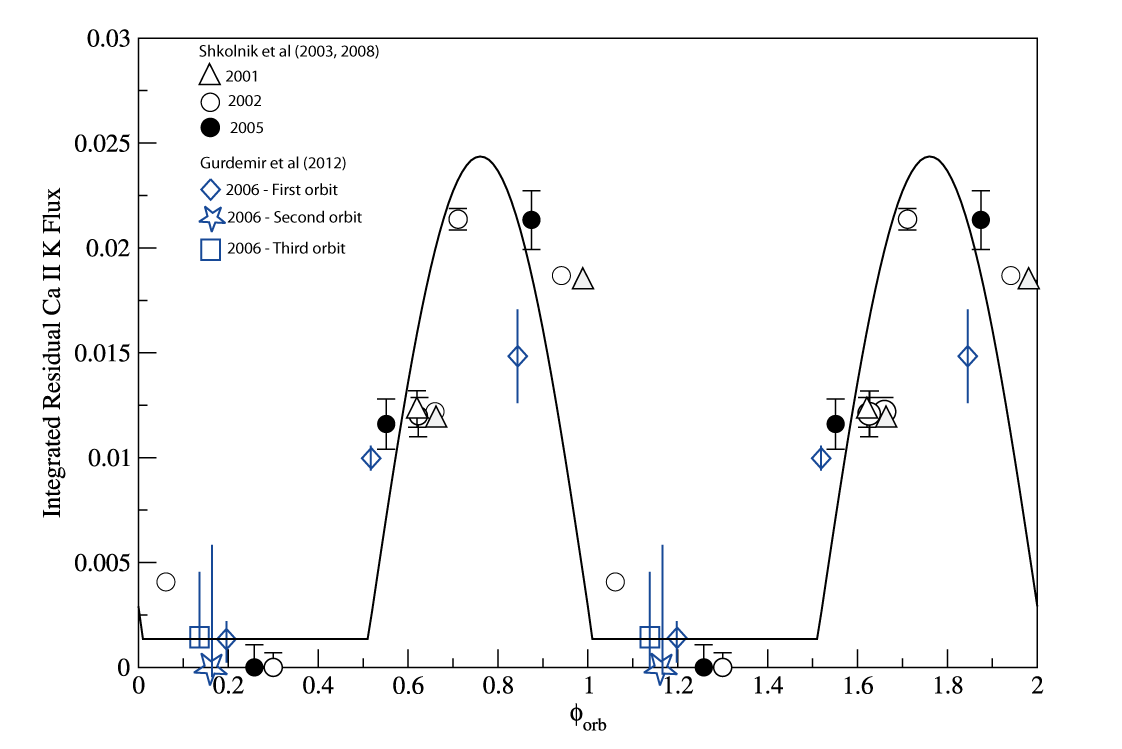}
\caption{Integrated Ca II K residual flux of HJ host HD 179949 as a function of orbital phase where $\phi$=0 is the sub-planetary point (or inferior conjunction).  The symbols are results from six individual epochs of observation collected from 2001 to 2006 by \cite{shkolnik2003,shkolnik2005,shkolnik2008} and \cite{gurdemir2012}.  The spot model shown is fit to the 2001-2005 data and shows a persistent region of excess chromospheric activity peaking at the planet's orbital phase of $\approx$0.75.
}
\label{fig:hd179949_spi}
\end{figure}

Although no tidally induced variability has yet been reported, magnetic SPI has seen a blossoming of data and modeling over the past 20 years. The strongest evidence for magnetic SPI is excess stellar activity modulated in phase with a planet as it orbits a star with a rotation period significantly different from the planet's orbital period. Such signatures were first reported by \cite{shkolnik2003} who observed periodic chromospheric activity through Ca II H \& K variability of HJ host HD 179949 modulated on the planet's orbital period of 3.092 d \citep{butler2006} rather than the stellar rotation period of 7 days \citep{fares2012}. Those data consisted of nightly high-resolution ($\lambda/\Delta\lambda\approx$110,000), high signal-to-noise (a few hundred per pixel) spectroscopy acquired over several epochs (Figure \ref{fig:hd179949_spi}, \citealt{shkolnik2005,shkolnik2008,gurdemir2012}).

In addition to HD 179949, several other stars with HJs exhibit this kind of Ca II H \& K modulation, including $\upsilon$ And, $\tau$ Boo, and HD 189733.  \cite{shkolnik2008} reported that this signature is present roughly $\approx$75\% of the time.  During other epochs only rotationally modulated spotting for these stars is observed. This is interpreted as variations in the stellar magnetic field configuration leading to weaker (or no) magnetic SPI with the planet's field. Simulations of magnetic SPI using magnetogram data of the varying solar magnetic fields confirm this to be a likely explanation of the intermittent effect \citep{cran07}. As another example, the large scale magnetic field of the planet-host star HD 189733 has been observed over multiple years using Zeeman-Doppler Imaging (ZDI) and the field shows structural evolution between observations \citep{moutou2007,fares2010,fares2013}. In this case, the SPI diagnostics in the HD 189733 system must vary with the stellar magnetic field strength and configuration.

\begin{figure}
\centering
\includegraphics[width=0.8\textwidth]{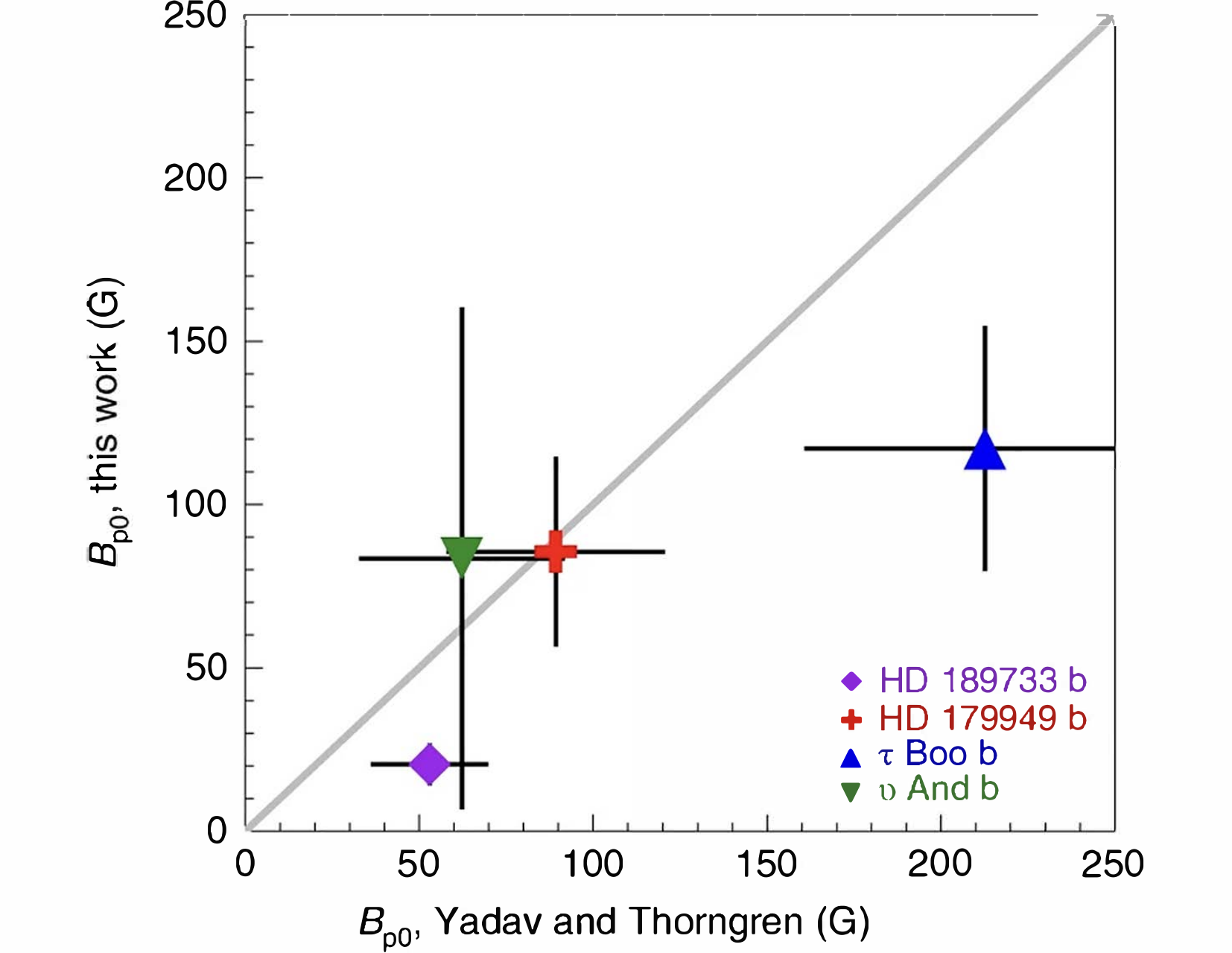}
\caption{Magnetic field strengths derived from Ca II K variability modulated by planet orbital phase (\citealt{Cauley2019}) compared to  those estimated through models of additional heat deposition (\citealt{yadav2017}) show a degree of similarity. The grey line is the 1-to-1 line and the error bars represent 1$\sigma$ uncertainties. This similarity between the magnetic field strengths from magnetic star-planet interactions and the predictions of the extra heat deposition models lends support to theories that attribute the generation of magnetic fields in hot planets to internal heat flux.}
\label{fig_cauley2019}       
\end{figure}

By estimating the energy emitted in the Calcium II K line, \cite{Cauley2018,Cauley2019} determined that the magnetic field strengths on the surfaces of four hot Jupiters vary between 20 G and 120 G (Figure~\ref{fig_cauley2019}). These figures are significantly higher—about 10 to 100 times—than what dynamo scaling laws would predict for planets with rotation periods between 2 to 4 days. Yet, these magnetic field strengths align with scaling laws that connect the strength of the magnetic field to the internal heat flow in giant planets (\citealt{yadav2017}). Such strong magnetic fields could enable the detection of electron cyclotron maser radio emissions, as they prevent the emissions from being absorbed by the planet's ionosphere. Close and continuous radio observations of hot Jupiter systems are essential to verify these magnetic field measurements and to understand how magnetic fields are formed in this significant group of exoplanets. Up to now, there has not been a definitive detection of radio emissions from exoplanets, which may be due to the lack of sensitivity for higher frequency emissions or the complex overlap with the radio signals from stars, as noted in studies by \cite{turner2021}, \cite{pineda2023}, and \cite{trigilio2023}. However, the reported observations of magnetic star-planet interactions in several star-exoplanet systems indicate that large exoplanets in close orbits do stimulate activity on the surfaces and in the atmospheres of their host stars, which can separately provide a way to investigate the magnetic fields of the exoplanets themselves.

\section{Scaling law to measure planetary magnetic field strengths}
\begin{figure}
\centering
\includegraphics[width=0.8\textwidth]{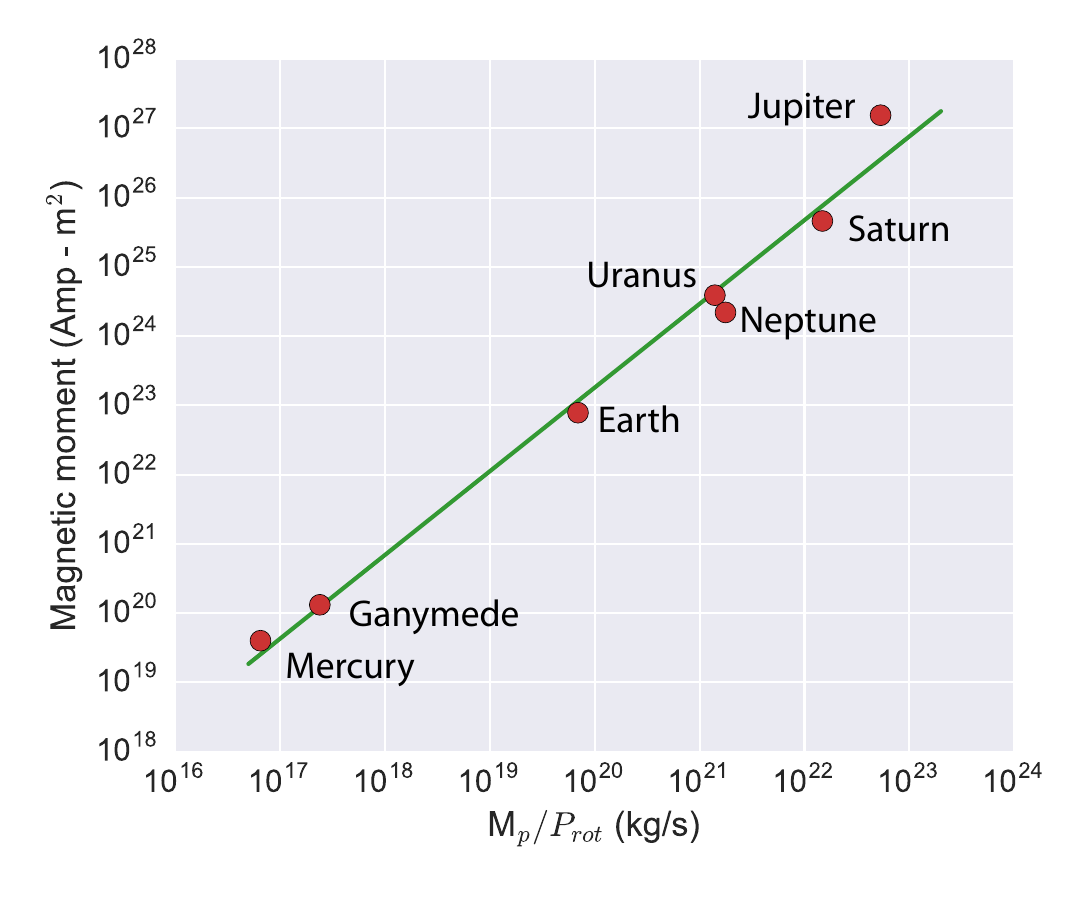}
\caption{The magnetic moment for the six magnetized solar system planets, plus Ganymede, plotted against the ratio of body mass to rotation period. The power law fit is $y\propto x^{1.21}$. Data are from Tholen et al. (2000) and Kivelson et al. (2002).}
\label{fig:magmom_ss}       
\end{figure}

\begin{figure}
\centering
\includegraphics[width=0.8\textwidth]{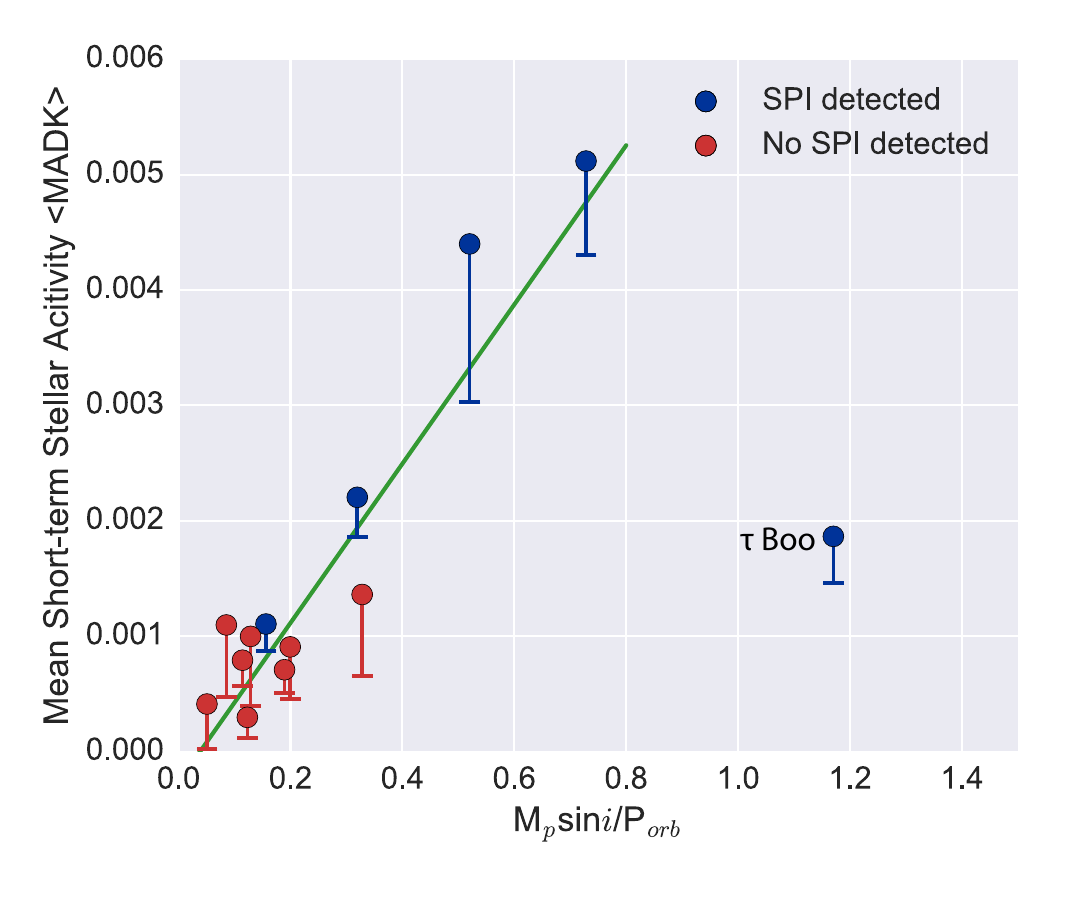}
\caption{$M_p\sin i/P_{\rm orb}$, which is proportional to the planet's magnetic moment (Figure~\ref{fig:magmom_ss}), plotted against the mean night-to-night Ca II K chromospheric activity (assuming the planet is tidally locked, $P_{\rm orb} = P_{{\rm rot},p}$). The blue symbols show systems where SPI has been detected. Note that  $\tau$ Boo, for which  $P_{\rm orb} = P_{{\rm rot},\star}$  does not follow the trend. This is evidence in support of a model (\citealt{lanza2009}) where near-zero relative motion of the planet through the stellar magnetosphere produces minimal magnetic SPI effects \citep{shkolnik2008}.}
\label{fig:magmom_exo}       
\end{figure}

In the solar system, there is a strong correlation between the magnetic moment of a body and the ratio of its mass to its rotation period (Figure \ref{fig:magmom_ss}).
Analogously, a similar relationship has emerged for exoplanets. Figure \ref{fig:magmom_exo} shows $M_p\sin i/P_{orb}$ against the stellar magnetic activity measure, $<$MADK$>$, the average of the Mean Absolute Deviation of Ca II K line variability per observing run. Note that the planet is assumed to be tidally locked such that $P_{\rm orb}$ equals the rotation period of the planet.

 There have been several attempts at providing an analytical prescription for magnetic SPI in the literature (e.g., \citealt{lanza2009,lanza2012,lanza2013}. Recently, \citet{Cauley2019} were able to constrain the potential models using observations. They showed that most promising scenario producing the necessary power for SPI is the Poynting flux across the base of a magnetic flux tube that connects the planetary surface to the stellar surface. This model was proposed by \citet{lanza2013}, and in this scenario, the available power can be expressed as:

\begin{equation}
    P_{\rm SPI} \approx \frac{2\pi}{\mu}f_{\rm AP}R_p^2B_{p0}^2v_{\rm rel},
\end{equation}
where $R_p$ is the radius of the planet, $B_{p0}$ is the magnetic field strength at the pole of the planet, $f_{\rm AP}$ is the fraction of the planetary hemisphere covered by flux tubes, $\mu$ is the magnetic permeability of free space, and $v_{\rm rel}$ is the relative velocity between the two bodies. This implies that systems with stars that are tidally locked to their HJs, i.e.,~stellar rotation period equals the orbital period as is the case for HJ host $\tau$~Boo, should produce weak P$_{SPI}$ (Figure \ref{fig:magmom_exo}; \citealt{shkolnik2008,walker2008,fares2013}).  

The strength of the planetary magnetic field for tidally locked planets has been a subject of debate but scaling laws presented by \cite{chri09} and others reviewed in \cite{chri10} predict that the planet's field strength depends primarily on the internal heat flux, and not on electrical conductivity nor the rotation speed. This same energy scaling can simultaneously explain the observed field strengths of Jupiter, Earth and rapidly rotating low-mass stars.

Using the formalism of \cite{lanza2013} above with the measured stellar magnetic fields from spectropolarimetric observation of these targets (e.g., \citealt{donati2008,fares2009,fares2013,jeffers2014,hussain2016,mengel2017}), it is possible to use this correlation to estimate the \textit{relative} magnetic field strengths of the planets, exhibiting  a range of planetary magnetic field strengths of the HJs in these systems. For example, the HJ around HD 179949 implies to have a field strength several times that of the HJ around HD 189733, consistent with the \cite{Cauley2019} measurement factor to be $\approx4\times$.

Since the planet-induced stellar activity can be more easily triggered by planets with larger magnetic moments, it is reasonable to assume that planets on eccentric orbits, which are are likely to spin more quickly compared to those in circular orbits, may be a fertile place to look for magnetic SPI. Although such efforts in radio have thus far come up empty-handed for an extremely eccentric planet (HD 80606 b; \citealt{lazio2010,degasperin2020}, the first tentative detection of magnetic SPI was recently reported in planet-phased photometry variability from TESS, NASA's Transiting Exoplanets Survey Satellite \citep{castro2024} on a somewhat evolved star HD~118203, which is orbited by an eccentric planet whose closest approach is as close as 0.086 AU.

There are stars for which no planet-phased activity is reported, e.g., HD 209458 \citep{shkolnik2008} and WASP 18 \citep{miller2015,pillitteri2014b}.  In these cases, the central stars are particularly inactive with very weak fields and measurable SPI is not expected according to Lanza's formalism as both the star and the planet require strong enough magnetic fields for an observable interaction.   In addition, in many cases, the data collected were of too low S/N to detect any induced modulations caused by the planet and/or lacked phase coverage of the planetary orbit making it difficult to disentangle planet-induced activity from stellar rotational modulation. The star may also have a highly variable magnetic field.  If the stellar magnetic field is highly complex in structure then it may be that the magnetic field lines simply do not reach the orbit of the planet \citep{lanza2009}. Finally, the planet itself may have a weak magnetic field or no field at all.

\section{Observations of planet induced variability at many wavelengths}
In addition to Ca II H \& K observations, planet phased modulation has been reported in broadband optical photometry from space for $\tau$ Boo \citep{walker2008} and CoRoT-2 \citep{pagano2009} and in X-ray for HD 17156 \citep{maggio2015}. Tentative evidence of planet-phased X-ray modulation of HD 179949 was reported by \cite{scandariato2013}. They find an activity modulation period of $\sim4$ days (with a false alarm probability of 2\%), which is longer than the orbital period of 3.1 days, but may be tracing the synodic period of the planet with respect to the star ($P_{syn}$=4.7--5.6 days for $P_{rot}$=7--9 days). Clearer planet-phased X-ray and far-UV modulation has also been reported for HD 189733 \citep{pillitteri2011,pillitteri2015}.

Ideally, simultaneous observations across optical, UV and X-ray activity indicators would be scheduled but has proven to be challenging to accomplish. From this and other perspectives discussed below, statistical studies of a large sample of monitored stars for planet phased stellar activity is the necessary path forward.

The HD 189733 system is one of the most studied as it is a bright K2V dwarf at a distance of 19.3 pc, hosts a transiting hot Jupiter at a distance of only 0.03 AU \citep{bouchy2005}, and exhibits planet induced Ca II H \& K variations \citep{shkolnik2005,shkolnik2008}. It has been the subject of multiple searches for X-ray flares that coincide with the orbit of the planet. Transit observations of HD 189733b, with phase coverage  from $\phi=0.52 - 0.65$ have shown that the X-ray spectrum softened in strict correspondence with the transit event, followed by a flaring event when the planet was at $\phi=0.54$ \citep{pillitteri2010,pillitteri2011,Pillitteri2014}. This phase offset for the beginning of the flare event corresponds to a location of $77^\circ$ forward of the sub-planetary point, as is also the case for the HD~179949 system. This phased emission is best interpreted as the observational signature of an active spot on the surface of the star that is connected to, and co-moving with, the planet \citep{Pillitteri2014}. Such a hot spot has been analytically derived by modeling the link between an exoplanet and the star \citep{lanza2012}. These authors calculated that if the planet is sufficiently close to the star (as is the case for hot Jupiters) the magnetic field lines that connect the star to the planet would produce such a phase offset owing to the relative orbital motion of the planet.

\section{Modeling SPI}
Simulations are also helping to understand SPI and planet phased emission through modeling studies aimed not at reproducing individual systems but rather at the general conditions that favor SPI. In the basic sense, SPI occurs when an obstacle (such as a magnetized planet) interacts with the magnetically driven thermal wind and embedded stellar wind.  

\begin{figure}
\centering
\includegraphics[width=0.8\textwidth]{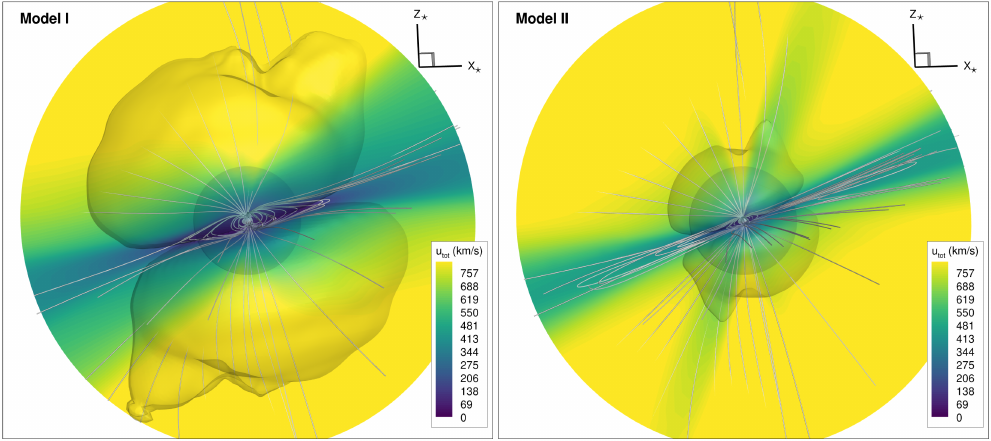}
\caption{3D MHD model of the stellar wind environment of GJ 436 using surface magnetic field maps obtained through Zeeman Doppler Imaging from \citet{bellotti2023}. The gray streamlines are the magnetic field lines embedded in the wind. The orbit of the planet is shown as a translucent sphere. The contour shows the wind velocity, and the Alfv\'en surface is shown by as a semi-opaque sphere. Figure from \citet{vidotto2023}}
\label{fig:gj436wind}       
\end{figure}

The stellar wind at the orbit of the planet is not a quantity that can be directly observed; however, using 3D MHD models that use a magnetic map of the stellar surface as input (e.g., through Zeeman Doppler Imaging; ZDI) the stellar wind can be extrapolated from the stellar surface to the orbit of the planet. Figure \ref{fig:gj436wind} shows the extrapolation of the stellar wind for the planet host GJ 436, an M dwarf, using an observed magnetic map obtained through ZDI as a boundary condition into the 3D MHD model BATS-R-US \citep{vidotto2023}. The simulation shows that at the orbital separation of the exoplanet (0.028 AU, $\sim$15$R_\star$) the planet is orbiting through both open and closed regions of stellar corona/wind. This variability of the field structure is expected to impact the observable signature of SPI, as was posited by \cite{shkolnik2008}. 

\citet{llama2013} carried out MHD modeling of HD 189733 using ZDI maps obtained one year apart by \citet{fares2010} as input to their model. Their study focused on the detectability and variability of a magnetospheric bowshock that will form when the stellar wind collides with the planetary field. They found that the size of the planetary magnetosphere expands and contracts by $\sim20\%$ as the planet orbits through various stellar wind conditions.

The first generation of SPI models focused primarily on recovering the phase offset between the sub-planetary point and the chromospheric hot spot rather than explaining the spot's energy dissipation \citep{mcivor2006,preusse2006,lanza2008}. The next generation of models explicitly included the planet and were able to show that the power generated in a reconnection event between the stellar corona and the planet can reproduce the observed hot spots \citep{lanza2009,cohen2011,lanza2013}.

An investigation by \citet{cohen2011} using the MHD code BATS-R-US showed that HD 189733b orbited in-and-out of the variable Alfv\'en radius and that when the planet was within the Alfv\'en radius its magnetosphere would reconnect with the stellar coronal field resulting in enhanced flaring from the host star. In their simulations the planet was implemented as an additional boundary condition representing HD 189733b's density, temperature, and magnetic field strength. They found that SPI varies during the planetary orbit and is highly dependent on the relative orientation of the stellar and planetary magnetic fields.

\begin{figure}[htb]
\centering
\includegraphics[width=\textwidth]{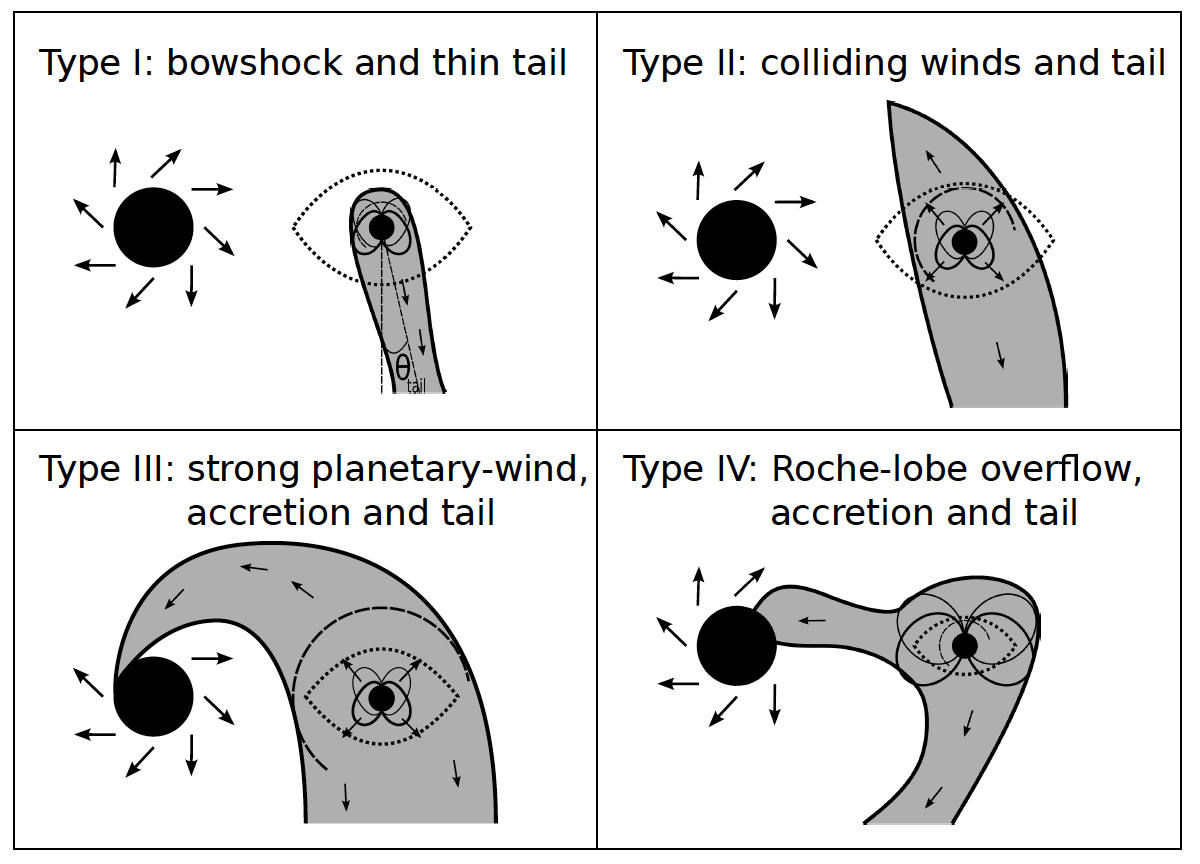}
\caption{The four types of star-planet interaction as described in \citet{matsakos2015}. In Type I, the ram and magnetic pressure of the stellar wind is greater than the planetary outflow, confining the material and leading to the formation of a bow shock (e.g., \citealt{vidotto2010,llama2011}). In Type II, the planetary outflow is stronger than in Type I resulting in material being swept back into a tail. The interactions of Types III and IV, the ram pressure of the stellar wind is greater than the tidal pressure of the planet, resulting in the formation of a tail behind the planet and an accretion stream onto the star. The accretion stream typically impacts the stellar surface $\sim90^\circ$ ahead of the sub-planetary point, in agreement with observations of magnetic SPI.}
\label{fig:matsakos}       
\end{figure}

A study by \citet{matsakos2015} was aimed at categorizing various types of SPI using the 3D MHD PLUTO code \citep{pluto2007,pluto2012}.  They ran 12 models in total, detailed in Table 2 of \cite{matsakos2015}. Since they were seeking to explore the parameter regime over which the observational signature of SPI changes, they chose to explore various parameters for the planet and star, rather than adopting the parameters for a known system. They classify star-planet interactions into four types illustrated in Figure \ref{fig:matsakos}. Types III and IV describe scenarios where an accretion stream forms between the planet and the star. For these interactions, the authors find that the ram pressure from the stellar wind must be greater than the magnetic and tidal pressures from the planet. The accretion stream arises through Kelvin-Helmholtz and Rayleigh-Taylor instabilities and is triggered by the interaction between the stellar wind and the denser planetary material. These simulations showed that the location where the accretion stream typically impacts the stellar surface is dependent on the parameters of the system but is typically $\sim45-90^\circ$ in front of the planet. This finding is in good agreement with the observed SPI phase offsets discussed above.

One of the current unknowns in modeling SPI is the conversion factor from the available flux for SPI into emission. The two competing theories are the Alfv\'en wing (AW) theory \citep{neubauer1998,zarka2001,saur2013} and the stretch-and-break (SB) theory \citep{lanza2013}. Simulations by \citealt{strugarek2016} were able to reproduce the conversion factor from the AW model and found that the process occurs before the SB scenario can be realized; however, observations by  \citet{Cauley2019} observed enhanced Ca II H\&K emission from the HD 189733 system, from which they were able to estimate the SPI power and found that it could only be reproduced using the higher power predicted by the stretch-and-break model.

\section{Statistical studies of magnetic SPI}

As the number of known exoplanets continuously rises, statistical studies are becoming an effective way to study the properties of exoplanetary systems. An efficient  strategy with which to study planet induced stellar emission is by analyzing single-epoch observations of a statistical sample in search of a significant difference in emission properties of stars with and without close-in giant planets.

From a sample of stars with Ca II H \& K observations, \cite{hart10} showed a correlation between planet surface gravities and the stellar log~R$^{\prime}_{HK}$ activity parameter for 23 systems with planets of M$_p$ $>$ 0.1 M$_J$ , $a$ $<$ 0.1 AU orbiting stars with 4200~K $<$ T$_{\rm eff} < $6200 K, with a weaker correlation with planet mass.  In another study of 210 systems, \cite{krej12} found statistically significant evidence that the equivalent width of the Ca II K line emission and log~R$^{\prime}_{HK}$ of the host star correlate with smaller semi-major axis and larger mass of the exoplanet, as would be expected for magnetic and tidal SPI.

The efficiency of extracting data from large photometric catalogs has made studying stellar activity of many more planet hosts possible in both the ultraviolet (UV) and X-ray.
A study of 72 exoplanet systems by \citet{poppenhaeger2010} showed no significant correlation between the fractional luminosity $(L_X/L_{\rm bol})$ with planet properties. They did, however, report a correlation of stellar X-ray luminosity with the ratio of planet mass to semi-major axis $(M_p\sin i/a)$, suggesting that massive, close-in planets tend to orbit more X-ray luminous stars.  They attributed this correlation to biases of
the radial velocity (RV) planet detection method, which favors smaller and further-out planets to be detected around less active, and thus X-ray faint, stars.

\begin{figure}
\centering
\includegraphics[width=0.8\textwidth]{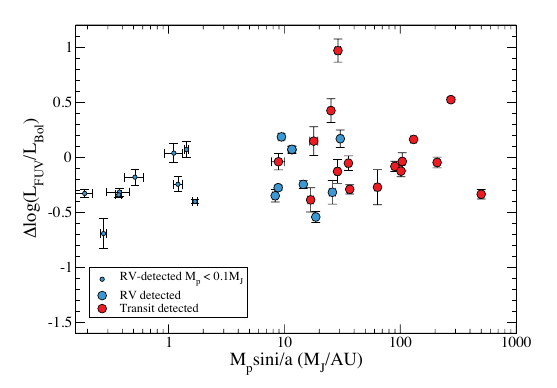}
\caption{The residual fractional FUV luminosity (i.e. photospheric flux removed leaving only stellar upper-atmospheric emission) as a function of the ratio of the planet mass to semi-major axis, a measure of star-planet interaction strength \citep{shko13}.}
\label{fig:shko13}       
\end{figure}

A study of both RV and transit detected planets by \cite{shko13} of the far-UV (FUV) emission as observed by the Galaxy Evolution Explorer (GALEX) also searched for evidence of increased stellar activity due to SPI in $\sim$300 FGK planet hosts.
This investigation found no clear correlations with $a$ or $M_p$,  yet reported tentative evidence for close-in massive planets (i.e.,~higher $M_p$/$a$) orbiting more FUV-active stars than those with far-out and/or smaller planets, in agreement with past X-ray and Ca II results (Figure~\ref{fig:shko13}). There may be less potential for detection bias in this case as transit-detected planets orbit stars with a more normal distribution of stellar activity than those with planets discovered with the RV method.  To confirm this, a sample of transiting small and distant planets still needs to be identified.

\begin{figure}
\centering
\includegraphics[width=0.8\textwidth]{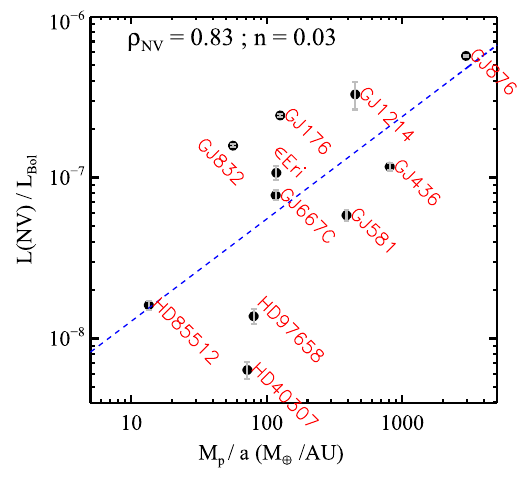}
\caption{Fractional N V (at 1240\AA) luminosity from a sample of 11 K and M dwarf planet hosts is weakly correlated with a measure of the star-planet interaction strength $M_p/a$, where $M_p$ is the mass of the most massive planet in the system (in Earth masses) and $a$ is the semi-major axis (in AU). The Pearson coefficient and statistical likelihood of a null correlation is shown at the top. This provides tentative evidence that the presence of short-period planets enhances the transition region activity on low-mass stars, possibly through the interaction of their magnetospheres (\citealt{fran16}). }
\label{fig:fran16}       
\end{figure}

The first statistical SPI test for lower mass (K and M) systems was reported by \cite{fran16} in which they measured a weak positive correlation between fractional N V luminosity, a transition region FUV emission line, with $M_p/a$ for the most massive planet in the system. They found tentative evidence that the presence of short-period planets (ranging in M$_p$sin$i$ from 3.5 to 615 M$_{Earth}$) enhances the transition region activity on low-mass stars, possibly through the interaction of their magnetospheres (Figure \ref{fig:fran16}).

\citet{cohen2015} modeled the interaction between an M-dwarf and a non-magnetized planet like Venus. Their work shows very different results for the localized space-weather environments for the planet for sub- and super-Alfv\'enic stellar wind conditions.
The authors postulate that these dynamic differences would lead to additional heating and additional energy being deposited into the atmosphere of the planet. In all their simulations, they find that the stellar wind penetrates much deeper into the atmosphere than for the magnetized planets simulated in \citet{cohen2014}, suggesting that for planets orbiting M dwarfs, a magnetosphere may be necessary to shield the planet's atmosphere.

\citet{vidotto2014} modeled the stellar wind of six M stars ranging from spectral type M0 to M2.5 to study the angular momentum of the host star and the rotational evolution of the star. They found the stellar wind to be highly structured at the orbital separation of the planet, and found that the planetary magnetospheric radii could vary by up to 20\% in a single orbit.
This will result in high variability in the strength of SPI signatures as the planet orbits through regions of closed and open magnetic field, implying that a larger, statistical study may be the most efficient path forward, especially for M dwarfs.

\section{Planetary Effects on Stellar Angular Momentum Evolution}

As the evidence continues to mount that star-planet interactions measurably increase stellar activity, and now for a wider range of planetary systems, there remains an ambiguity in the larger statistical, single-epoch studies as to whether or not this effect is caused by magnetic SPI, tidal SPI or planet search selection biases. Although no tidal SPI has been observed as stellar activity modulated by half the planet's orbital period (\citealt{cuntz2000}), there may be other effects due to the presence of the planets or planet formation process on the angular momentum evolution of the stars, which might increase the stellar rotation through tidal spin-up or decrease the efficiency of stellar magnetic breaking \citep{lanza2010b,cohen2011}. In both cases, the star would be more active than expected for its mass and age.  

For main-sequence FGK stars, the magnetized stellar wind acts as a brake on the stellar rotation, decreasing the global stellar activity rate as the star ages. This well-observed process has led to the so-called ``age-rotation-activity relationship ''. However, the presence of a short-period giant planet may affect the star's angular momentum. Under this scenario, the age-activity relation will systematically underestimate the star's age, potentially making ``gyrochronology'' inapplicable to these systems. This poses an issue for evolutionary studies of exoplanets and their host stars, including planet migration models and planet atmospheric evolution. In a possibly common case, when giant planet engulfment occurs as as result of high-eccentricity planetary migration, conservation of energy and momentum calculations can explain the observed increased stellar spin rates in young clusters, hinting that planets falling into their stars might be a common part of how planetary systems evolve (\citealt{Qureshi2018}).

Several studies have found that stars hosting giant planets rotate faster than the evolutionary models predict. This increase in rotation rate is thought to be the direct consequence of tidal spin-up of the star by the planet. Additional evidence for the tidal spin-up of stars by giant planets has been found using two hot Jupiter systems by \citet{schro2011} and \citet{pillitteri2011}. These studies searched for X-ray emission from M dwarf companions to the active planet hosts CoRoT-2 and HD 189733. Both systems did not show X-ray emission, indicating that the age of the systems to be $> 2$Gyr; however, the rotation-age relation places these systems between 100-300 Myr for CoRoT-2 and 600 Myr for HD 189733.

A study by \citet{lanza2010b} showed that tides alone cannot spin-up the star to the levels seen in CoRoT-2 and HD 189733. Rather, his study postulated that the excess rotation is a consequence of interactions between the planetary magnetic field and the stellar coronal field. He proposed that these interactions would result in a magnetic field topology in which the majority of the field lines are closed. This configuration would therefore limit the efficiency of the stellar wind to spin-down the star through angular momentum loss. By computing a simple linear force-free model, Lanza (2010) was able to compute the radial extension of the stellar corona and its angular momentum loss. He found that stars that host hot Jupiters show a much slower angular momentum loss rate than similar stars without a short-period giant planet, similar to \cite{cohen2011}.

In order to disentangle the possible causes of the observed increased stellar activity of HJ hosts observed from single-epoch observations, it is necessary to monitor the activity throughout the planet's orbit and over the stellar rotation period.   Such studies can better characterize the star's variability, generate firmer statistical results of any planet-induced activity, and assess the underlining physical processes involved.

\section{Summary}
Detecting exoplanetary magnetic fields enables us to probe the internal structures of the planets and to place better constraints on their atmospheric mass loss through erosion from the stellar wind. Searching for the observational signatures of magnetic SPI in the form of planet-induced stellar activity has proved to be the most successful method to date for detecting magnetic fields of hot Jupiters. 

Single-epoch statistical studies in search SPI signatures show that indeed there are significant differences in the activity levels between stars with close-in giant planets compared to those without them. However, the cause of this remains ambiguous with four possible explanations.  
\begin{itemize}
    \item Induced stellar activity in the form of interactions between the stellar and planetary magnetic fields.
    \item The inhibition of magnetic breaking and thus faster than expected stellar rotation and increased stellar activity.
    \item Tidal spin-up of the star due the presence of the close-in planet.
    \item Lastly, the selection biases of planet-hunting techniques.  
\end{itemize}

These potential underlying causes of such a result highlight the need for further monitoring campaigns across planetary orbit and stellar rotation periods to clearly identify planet-induced excess stellar activity.  

The vast majority of SPI studies, both individual monitoring as well as larger single-epoch statistical studies, have concentrated on main-sequence FGK stars as they are the dominant hosts of hot Jupiters. 
These stars have the advantage of being relatively quiescent compared to M dwarfs, and thus teasing out signals produced by magnetic SPI from intrinsic stellar activity is simpler. But they also have the disadvantage of lower stellar magnetic field strengths compared to M dwarfs, lowering the power produced by the interaction. 

The modeling of magnetic SPI, especially with realistic stellar magnetic maps from ZDI surveys, continues to advance and aid in the interpretation of observed planet-phased enhanced activity across the main sequence. Additional models enable quantitative predictions of the radio flux density  for stars displaying signatures of SPI. Radio detections of at least a few of these systems will help calibrate the relative field strengths, and provide for the first time, true magnetic field strengths for hot Jupiters.

Ongoing and future studies of magnetic SPI in a large sample of systems are necessary for improved statistics and distributions of magnetic fields of exoplanets. 
Extensions of these techniques to other tightly orbiting stellar systems, such as smaller planets close to M dwarfs, are challenging but possible. In these systems, star-planet separations of tens of stellar radii begin to coincide with the radiative habitable zone where planetary magnetic fields are likely a necessary condition for surface habitability.
As more close-in planets around relatively bright M dwarfs are discovered, the search for magnetic star-planet interactions will be extended to these low-mass stars.

\section{Cross-References}
\begin{itemize}
\item{Stellar activity and its impact on planets}
\item{Magnetic fields in planet hosting stars}
\item{Star-planet interactions in the radio domain: Prospect for their detection}
\item{Models of star-planet magnetic interaction}
\item{Stellar coronal and wind models: Impact on exoplanets}
\item{Accretion of planetary material onto their host stars}
\item{Planetary habitability and magnetic fields}
\item{Star-planet interactions and habitability: Radiative effects}
\item{Star-planet interactions and habitability: Gravitational effects}
\end{itemize}

\bibliographystyle{spbasicHBexo}  
\bibliography{chapter_revision_refs_2024} 

\begin{thebibliography}{89}
\providecommand{\natexlab}[1]{#1}
\providecommand{\url}[1]{{#1}}
\providecommand{\urlprefix}{URL }
\expandafter\ifx\csname urlstyle\endcsname\relax
  \providecommand{\doi}[1]{DOI~\discretionary{}{}{}#1}\else
  \providecommand{\doi}{DOI~\discretionary{}{}{}\begingroup \urlstyle{rm}\Url}\fi
\providecommand{\eprint}[2][]{\url{#2}}

\bibitem[{{Bastian} et~al.(2000){Bastian}, {Dulk}, and {Leblanc}}]{bastian2000}
{Bastian} TS, {Dulk} GA {Leblanc} Y (2000) {A Search for Radio Emission from Extrasolar Planets}. \apj 545:1058--1063

\bibitem[{{Bellotti} et~al.(2023){Bellotti}, {Fares}, {Vidotto}, {Morin}, {Petit}, {Hussain}, {Bourrier}, {Donati}, {Moutou}, and {H{\'e}brard}}]{bellotti2023}
{Bellotti} S, {Fares} R, {Vidotto} AA et~al. (2023) {The space weather around the exoplanet GJ 436b. I. The large-scale stellar magnetic field}. \aap 676:A139

\bibitem[{{Bouchy} et~al.(2005){Bouchy}, {Udry}, {Mayor}, {Moutou}, {Pont}, {Iribarne}, {da Silva}, {Ilovaisky}, {Queloz}, {Santos}, {S{\'e}gransan}, and {Zucker}}]{bouchy2005}
{Bouchy} F, {Udry} S, {Mayor} M et~al. (2005) {ELODIE metallicity-biased search for transiting Hot Jupiters. II. A very hot Jupiter transiting the bright K star HD 189733}. \aap 444:L15--L19

\bibitem[{{Butler} et~al.(2006){Butler}, {Wright}, {Marcy}, {Fischer}, {Vogt}, {Tinney}, {Jones}, {Carter}, {Johnson}, {McCarthy}, and {Penny}}]{butler2006}
{Butler} RP, {Wright} JT, {Marcy} GW et~al. (2006) {Catalog of Nearby Exoplanets}. \apj 646:505--522

\bibitem[{{Castro-Gonz{\'a}lez} et~al.(2024){Castro-Gonz{\'a}lez}, {Lillo-Box}, {Correia}, {Santos}, {Barrado}, {Morales-Calder{\'o}n}, and {Shkolnik}}]{castro2024}
{Castro-Gonz{\'a}lez} A, {Lillo-Box} J, {Correia} ACM et~al. (2024) {Signs of magnetic star-planet interactions in HD 118203. TESS detects stellar variability that matches the orbital period of a close-in eccentric Jupiter-sized companion}. arXiv e-prints arXiv:2401.17272

\bibitem[{Cauley et~al.(2018)Cauley, Shkolnik, Llama, Bourrier, and Moutou}]{Cauley2018}
Cauley PW, Shkolnik EL, Llama J, Bourrier V Moutou C (2018) Evidence of {{Magnetic Star-Planet Interactions}} in the {{HD}} 189733 {{System}} from {{Orbitally Phased Ca II K Variations}}. The Astronomical Journal 156:262

\bibitem[{{Cauley} et~al.(2019){Cauley}, {Shkolnik}, {Llama}, and {Lanza}}]{Cauley2019}
{Cauley} PW, {Shkolnik} EL, {Llama} J {Lanza} AF (2019) {Magnetic field strengths of hot Jupiters from signals of star-planet interactions}. Nature Astronomy 3:1128--1134

\bibitem[{{Christensen}(2010)}]{chri10}
{Christensen} UR (2010) {Dynamo Scaling Laws and Applications to the Planets}. \ssr 152:565--590

\bibitem[{{Christensen} et~al.(2009){Christensen}, {Holzwarth}, and {Reiners}}]{chri09}
{Christensen} UR, {Holzwarth} V {Reiners} A (2009) {Energy flux determines magnetic field strength of planets and stars}. \nat 457:167--169

\bibitem[{{Cohen} et~al.(2009){Cohen}, {Drake}, {Kashyap}, {Saar}, {Sokolov}, {Manchester}, {Hansen}, and {Gombosi}}]{cohen2009}
{Cohen} O, {Drake} JJ, {Kashyap} VL et~al. (2009) {Interactions of the Magnetospheres of Stars and Close-In Giant Planets}. \apjl 704:L85--L88

\bibitem[{{Cohen} et~al.(2011){Cohen}, {Kashyap}, {Drake}, {Sokolov}, {Garraffo}, and {Gombosi}}]{cohen2011}
{Cohen} O, {Kashyap} VL, {Drake} JJ et~al. (2011) {The Dynamics of Stellar Coronae Harboring Hot Jupiters. I. A Time-dependent Magnetohydrodynamic Simulation of the Interplanetary Environment in the HD 189733 Planetary System}. \apj 733:67

\bibitem[{{Cohen} et~al.(2014){Cohen}, {Drake}, {Glocer}, {Garraffo}, {Poppenhaeger}, {Bell}, {Ridley}, and {Gombosi}}]{cohen2014}
{Cohen} O, {Drake} JJ, {Glocer} A et~al. (2014) {Magnetospheric Structure and Atmospheric Joule Heating of Habitable Planets Orbiting M-dwarf Stars}. \apj 790:57

\bibitem[{{Cohen} et~al.(2015){Cohen}, {Ma}, {Drake}, {Glocer}, {Garraffo}, {Bell}, and {Gombosi}}]{cohen2015}
{Cohen} O, {Ma} Y, {Drake} JJ et~al. (2015) {The Interaction of Venus-like, M-dwarf Planets with the Stellar Wind of Their Host Star}. \apj 806:41

\bibitem[{{Cranmer} and {Saar}(2007)}]{cran07}
{Cranmer} SR {Saar} SH (2007) {Exoplanet-Induced Chromospheric Activity: Realistic Light Curves from Solar-type Magnetic Fields}. ArXiv Astrophysics e-prints

\bibitem[{{Cuntz} et~al.(2000){Cuntz}, {Saar}, and {Musielak}}]{cuntz2000}
{Cuntz} M, {Saar} SH {Musielak} ZE (2000) {On Stellar Activity Enhancement Due to Interactions with Extrasolar Giant Planets}. \apjl 533:L151--L154

\bibitem[{{de Gasperin} et~al.(2020){de Gasperin}, {Lazio}, and {Knapp}}]{degasperin2020}
{de Gasperin} F, {Lazio} TJW {Knapp} M (2020) {Radio observations of HD 80606 near planetary periastron. II. LOFAR low band antenna observations at 30-78 MHz}. \aap 644:A157

\bibitem[{{do Nascimento} et~al.(2016){do Nascimento}, {Vidotto}, {Petit}, {Folsom}, {Castro}, {Marsden}, {Morin}, {Porto de Mello}, {Meibom}, {Jeffers}, {Guinan}, and {Ribas}}]{doNascimento}
{do Nascimento} JD Jr, {Vidotto} AA, {Petit} P et~al. (2016) {Magnetic Field and Wind of Kappa Ceti: Toward the Planetary Habitability of the Young Sun When Life Arose on Earth}. \apjl 820:L15

\bibitem[{{Donati} et~al.(2008){Donati}, {Moutou}, {Far{\`e}s}, {Bohlender}, {Catala}, {Deleuil}, {Shkolnik}, {Collier Cameron}, {Jardine}, and {Walker}}]{donati2008}
{Donati} JF, {Moutou} C, {Far{\`e}s} R et~al. (2008) {Magnetic cycles of the planet-hosting star {$\tau$} Bootis}. \mnras 385:1179--1185

\bibitem[{{Dressing} and {Charbonneau}(2015)}]{dres15}
{Dressing} CD {Charbonneau} D (2015) {The Occurrence of Potentially Habitable Planets Orbiting M Dwarfs Estimated from the Full Kepler Dataset and an Empirical Measurement of the Detection Sensitivity}. \apj 807:45

\bibitem[{{Fares} et~al.(2009){Fares}, {Donati}, {Moutou}, {Bohlender}, {Catala}, {Deleuil}, {Shkolnik}, {Collier Cameron}, {Jardine}, and {Walker}}]{fares2009}
{Fares} R, {Donati} JF, {Moutou} C et~al. (2009) {Magnetic cycles of the planet-hosting star {$\tau$} Bootis - II. A second magnetic polarity reversal}. \mnras 398:1383--1391

\bibitem[{{Fares} et~al.(2010){Fares}, {Donati}, {Moutou}, {Jardine}, {Grie{\ss}meier}, {Zarka}, {Shkolnik}, {Bohlender}, {Catala}, and {Collier Cameron}}]{fares2010}
{Fares} R, {Donati} JF, {Moutou} C et~al. (2010) {Searching for star-planet interactions within the magnetosphere of HD189733}. \mnras 406:409--419

\bibitem[{{Fares} et~al.(2012){Fares}, {Donati}, {Moutou}, {Jardine}, {Cameron}, {Lanza}, {Bohlender}, {Dieters}, {Mart{\'{\i}}nez Fiorenzano}, {Maggio}, {Pagano}, and {Shkolnik}}]{fares2012}
{Fares} R, {Donati} JF, {Moutou} C et~al. (2012) {Magnetic field, differential rotation and activity of the hot-Jupiter-hosting star HD 179949}. \mnras 423:1006--1017

\bibitem[{{Fares} et~al.(2013){Fares}, {Moutou}, {Donati}, {Catala}, {Shkolnik}, {Jardine}, {Cameron}, and {Deleuil}}]{fares2013}
{Fares} R, {Moutou} C, {Donati} JF et~al. (2013) {A small survey of the magnetic fields of planet-host stars}. \mnras 435:1451--1462

\bibitem[{{Farrell} et~al.(1999){Farrell}, {Desch}, and {Zarka}}]{farrell1999}
{Farrell} WM, {Desch} MD {Zarka} P (1999) {On the possibility of coherent cyclotron emission from extrasolar planets}. \jgr 104:14,025--14,032

\bibitem[{{France} et~al.(2016){France}, {Parke Loyd}, {Youngblood}, {Brown}, {Schneider}, {Hawley}, {Froning}, {Linsky}, {Roberge}, {Buccino}, {Davenport}, {Fontenla}, {Kaltenegger}, {Kowalski}, {Mauas}, {Miguel}, {Redfield}, {Rugheimer}, {Tian}, {Vieytes}, {Walkowicz}, and {Weisenburger}}]{fran16}
{France} K, {Parke Loyd} RO, {Youngblood} A et~al. (2016) {The MUSCLES Treasury Survey. I. Motivation and Overview}. \apj 820:89

\bibitem[{{Gurdemir} et~al.(2012){Gurdemir}, {Redfield}, and {Cuntz}}]{gurdemir2012}
{Gurdemir} L, {Redfield} S {Cuntz} M (2012) {Planet-Induced Emission Enhancements in HD 179949: Results from McDonald Observations}. \pasa 29:141--149

\bibitem[{{Hartman}(2010)}]{hart10}
{Hartman} JD (2010) {A Correlation Between Stellar Activity and the Surface Gravity of Hot Jupiters}. \apjl 717:L138--L142

\bibitem[{{Hussain} et~al.(2016){Hussain}, {Alvarado-G{\'o}mez}, {Grunhut}, {Donati}, {Alecian}, {Oksala}, {Morin}, {Fares}, {Jardine}, {Drake}, {Cohen}, {Matt}, {Petit}, {Redfield}, and {Walter}}]{hussain2016}
{Hussain} GAJ, {Alvarado-G{\'o}mez} JD, {Grunhut} J et~al. (2016) {A spectro-polarimetric study of the planet-hosting G dwarf, HD 147513}. \aap 585:A77

\bibitem[{{Jardine} and {Collier Cameron}(2008)}]{jardine2008}
{Jardine} M {Collier Cameron} A (2008) {Radio emission from exoplanets: the role of the stellar coronal density and magnetic field strength}. \aap 490:843--851

\bibitem[{{Jeffers} et~al.(2014){Jeffers}, {Petit}, {Marsden}, {Morin}, {Donati}, and {Folsom}}]{jeffers2014}
{Jeffers} SV, {Petit} P, {Marsden} SC et~al. (2014) {{$\epsilon$} Eridani: an active K dwarf and a planet hosting star?. The variability of its large-scale magnetic field topology}. \aap 569:A79

\bibitem[{{Kavanagh} and {Vedantham}(2023)}]{kavanagh2023}
{Kavanagh} RD {Vedantham} HK (2023) {Hunting for exoplanets via magnetic star-planet interactions: geometrical considerations for radio emission}. \mnras 524(4):6267--6284

\bibitem[{{Krej{\v c}ov{\'a}} and {Budaj}(2012)}]{krej12}
{Krej{\v c}ov{\'a}} T {Budaj} J (2012) {Evidence for enhanced chromospheric Ca II H and K emission in stars with close-in extrasolar planets}. \aap 540:A82

\bibitem[{{Lanza}(2008)}]{lanza2008}
{Lanza} AF (2008) {Hot Jupiters and stellar magnetic activity}. \aap 487:1163--1170

\bibitem[{{Lanza}(2009)}]{lanza2009}
{Lanza} AF (2009) {Stellar coronal magnetic fields and star-planet interaction}. \aap 505:339--350

\bibitem[{{Lanza}(2010)}]{lanza2010b}
{Lanza} AF (2010) {Hot Jupiters and the evolution of stellar angular momentum}. \aap 512:A77

\bibitem[{{Lanza}(2012)}]{lanza2012}
{Lanza} AF (2012) {Star-planet magnetic interaction and activity in late-type stars with close-in planets}. \aap 544:A23

\bibitem[{{Lanza}(2013)}]{lanza2013}
{Lanza} AF (2013) {Star-planet magnetic interaction and evaporation of planetary atmospheres}. \aap 557:A31

\bibitem[{{Lazio} et~al.(2009){Lazio}, {Bastian}, {Bryden}, {Farrell}, {Griessmeier}, {Hallinan}, {Kasper}, {Kuiper}, {Lecacheux}, {Majid}, {Osten}, {Shklonik}, {Stevens}, {Winterhalter}, and {Zarka}}]{lazio2009}
{Lazio} J, {Bastian} T, {Bryden} G et~al. (2009) {Magnetospheric Emissions from Extrasolar Planets}. In: astro2010: The Astronomy and Astrophysics Decadal Survey, Astronomy, vol 2010

\bibitem[{{Lazio} et~al.(2010){Lazio}, {Shankland}, {Farrell}, and {Blank}}]{lazio2010}
{Lazio} TJW, {Shankland} PD, {Farrell} WM {Blank} DL (2010) {Radio Observations of HD 80606 Near Planetary Periastron}. \aj 140(6):1929--1933

\bibitem[{{Lazio} et~al.(2016){Lazio}, {Shkolnik}, {Hallinan}, and {Planetary Habitability Study Team}}]{lazio2016}
{Lazio} TJW, {Shkolnik} E, {Hallinan} G {Planetary Habitability Study Team} (2016) {Planetary Magnetic Fields: Planetary Interiors and Habitability}. Tech. rep.

\bibitem[{{Llama} et~al.(2011){Llama}, {Wood}, {Jardine}, {Vidotto}, {Helling}, {Fossati}, and {Haswell}}]{llama2011}
{Llama} J, {Wood} K, {Jardine} M et~al. (2011) {The shocking transit of WASP-12b: modelling the observed early ingress in the near-ultraviolet}. \mnras 416:L41--L44

\bibitem[{{Llama} et~al.(2013){Llama}, {Vidotto}, {Jardine}, {Wood}, {Fares}, and {Gombosi}}]{llama2013}
{Llama} J, {Vidotto} AA, {Jardine} M et~al. (2013) {Exoplanet transit variability: bow shocks and winds around HD 189733b}. \mnras 436:2179--2187

\bibitem[{{Maggio} et~al.(2015){Maggio}, {Pillitteri}, {Scandariato}, {Lanza}, {Sciortino}, {Borsa}, {Bonomo}, {Claudi}, {Covino}, {Desidera}, {Gratton}, {Micela}, {Pagano}, {Piotto}, {Sozzetti}, {Cosentino}, and {Maldonado}}]{maggio2015}
{Maggio} A, {Pillitteri} I, {Scandariato} G et~al. (2015) {Coordinated X-Ray and Optical Observations of Star-Planet Interaction in HD 17156}. \apjl 811:L2

\bibitem[{{Matsakos} et~al.(2015){Matsakos}, {Uribe}, and {K{\"o}nigl}}]{matsakos2015}
{Matsakos} T, {Uribe} A {K{\"o}nigl} A (2015) {Classification of magnetized star-planet interactions: bow shocks, tails, and inspiraling flows}. \aap 578:A6

\bibitem[{{McIvor} et~al.(2006){McIvor}, {Jardine}, and {Holzwarth}}]{mcivor2006}
{McIvor} T, {Jardine} M {Holzwarth} V (2006) {Extrasolar planets, stellar winds and chromospheric hotspots}. \mnras 367:L1--L5

\bibitem[{{Mengel} et~al.(2017){Mengel}, {Marsden}, {Carter}, {Horner}, {King}, {Fares}, {Jeffers}, {Petit}, {Vidotto}, {Morin}, and {BCool Collaboration}}]{mengel2017}
{Mengel} MW, {Marsden} SC, {Carter} BD et~al. (2017) {A BCool survey of the magnetic fields of planet-hosting solar-type stars}. \mnras 465:2734--2747

\bibitem[{{Mignone} et~al.(2007){Mignone}, {Bodo}, {Massaglia}, {Matsakos}, {Tesileanu}, {Zanni}, and {Ferrari}}]{pluto2007}
{Mignone} A, {Bodo} G, {Massaglia} S et~al. (2007) {PLUTO: A Numerical Code for Computational Astrophysics}. \apjs 170:228--242

\bibitem[{{Mignone} et~al.(2012){Mignone}, {Zanni}, {Tzeferacos}, {van Straalen}, {Colella}, and {Bodo}}]{pluto2012}
{Mignone} A, {Zanni} C, {Tzeferacos} P et~al. (2012) {The PLUTO Code for Adaptive Mesh Computations in Astrophysical Fluid Dynamics}. \apjs 198:7

\bibitem[{{Miller} et~al.(2015){Miller}, {Gallo}, {Wright}, and {Pearson}}]{miller2015}
{Miller} BP, {Gallo} E, {Wright} JT {Pearson} EG (2015) {A Comprehensive Statistical Assessment of Star-Planet Interaction}. \apj 799:163

\bibitem[{{Moutou} et~al.(2007){Moutou}, {Donati}, {Savalle}, {Hussain}, {Alecian}, {Bouchy}, {Catala}, {Collier Cameron}, {Udry}, and {Vidal-Madjar}}]{moutou2007}
{Moutou} C, {Donati} JF, {Savalle} R et~al. (2007) {Spectropolarimetric observations of the transiting planetary system of the K dwarf HD 189733}. \aap 473:651--660

\bibitem[{{Neubauer}(1998)}]{neubauer1998}
{Neubauer} FM (1998) {The sub-Alfv{\'e}nic interaction of the Galilean satellites with the Jovian magnetosphere}. \jgr 103(E9):19,843--19,866

\bibitem[{{Pagano} et~al.(2009){Pagano}, {Lanza}, {Leto}, {Messina}, {Barge}, and {Baglin}}]{pagano2009}
{Pagano} I, {Lanza} AF, {Leto} G et~al. (2009) {CoRoT-2a Magnetic Activity: Hints for Possible Star-Planet Interaction}. Earth Moon and Planets 105:373--378

\bibitem[{{Pillitteri} et~al.(2010){Pillitteri}, {Wolk}, {Cohen}, {Kashyap}, {Knutson}, {Lisse}, and {Henry}}]{pillitteri2010}
{Pillitteri} I, {Wolk} SJ, {Cohen} O et~al. (2010) {XMM-Newton Observations of HD 189733 During Planetary Transits}. \apj 722:1216--1225

\bibitem[{{Pillitteri} et~al.(2011){Pillitteri}, {G{\"u}nther}, {Wolk}, {Kashyap}, and {Cohen}}]{pillitteri2011}
{Pillitteri} I, {G{\"u}nther} HM, {Wolk} SJ, {Kashyap} VL {Cohen} O (2011) {X-Ray Activity Phased with Planet Motion in HD 189733?} \apjl 741:L18

\bibitem[{{Pillitteri} et~al.(2014{\natexlab{a}}){Pillitteri}, {Wolk}, {Lopez-Santiago}, {G{\"u}nther}, {Sciortino}, {Cohen}, {Kashyap}, and {Drake}}]{Pillitteri2014}
{Pillitteri} I, {Wolk} SJ, {Lopez-Santiago} J et~al. (2014{\natexlab{a}}) {The Corona of HD 189733 and its X-Ray Activity}. \apj 785:145

\bibitem[{{Pillitteri} et~al.(2014{\natexlab{b}}){Pillitteri}, {Wolk}, {Sciortino}, and {Antoci}}]{pillitteri2014b}
{Pillitteri} I, {Wolk} SJ, {Sciortino} S {Antoci} V (2014{\natexlab{b}}) {No X-rays from WASP-18. Implications for its age, activity, and the influence of its massive hot Jupiter}. \aap 567:A128

\bibitem[{{Pillitteri} et~al.(2015){Pillitteri}, {Maggio}, {Micela}, {Sciortino}, {Wolk}, and {Matsakos}}]{pillitteri2015}
{Pillitteri} I, {Maggio} A, {Micela} G et~al. (2015) {FUV Variability of HD 189733. Is the Star Accreting Material From Its Hot Jupiter?} \apj 805:52

\bibitem[{Pineda and Villadsen(2023)}]{pineda2023}
Pineda JS Villadsen J (2023) Coherent radio bursts from known {{M-dwarf}} planet-host {{YZ Ceti}}. Nature Astronomy 7(5):569--578

\bibitem[{{Piskunov}(1996)}]{piskunov1996}
{Piskunov} N (1996) {Doppler imaging of eclipsing binaries}. In: {Strassmeier} KG {Linsky} JL (eds) Stellar Surface Structure, IAU Symposium, vol 176, p~45

\bibitem[{{Poppenhaeger} et~al.(2010){Poppenhaeger}, {Robrade}, and {Schmitt}}]{poppenhaeger2010}
{Poppenhaeger} K, {Robrade} J {Schmitt} JHMM (2010) {Coronal properties of planet-bearing stars}. \aap 515:A98

\bibitem[{{Powell} et~al.(1999){Powell}, {Roe}, {Linde}, {Gombosi}, and {De Zeeuw}}]{powell1999}
{Powell} KG, {Roe} PL, {Linde} TJ, {Gombosi} TI {De Zeeuw} DL (1999) {A Solution-Adaptive Upwind Scheme for Ideal Magnetohydrodynamics}. Journal of Computational Physics 154:284--309

\bibitem[{{Preusse} et~al.(2006){Preusse}, {Kopp}, {B{\"u}chner}, and {Motschmann}}]{preusse2006}
{Preusse} S, {Kopp} A, {B{\"u}chner} J {Motschmann} U (2006) {A magnetic communication scenario for hot Jupiters}. \aap 460:317--322

\bibitem[{{Qureshi} et~al.(2018){Qureshi}, {Naoz}, and {Shkolnik}}]{Qureshi2018}
{Qureshi} A, {Naoz} S {Shkolnik} EL (2018) {Signature of Planetary Mergers on Stellar Spins}. \apj 864(1):65

\bibitem[{{Saar} et~al.(2004){Saar}, {Cuntz}, and {Shkolnik}}]{saar2004}
{Saar} SH, {Cuntz} M {Shkolnik} E (2004) {Stellar Activity Enhancement by Planets: Theory and Observations}. In: {Dupree} AK {Benz} AO (eds) Stars as Suns : Activity, Evolution and Planets, IAU Symposium, vol 219, p 355

\bibitem[{{Saur} et~al.(2013){Saur}, {Grambusch}, {Duling}, {Neubauer}, and {Simon}}]{saur2013}
{Saur} J, {Grambusch} T, {Duling} S, {Neubauer} FM {Simon} S (2013) {Magnetic energy fluxes in sub-Alfv{\'e}nic planet star and moon planet interactions}. \aap 552:A119

\bibitem[{{Scandariato} et~al.(2013){Scandariato}, {Maggio}, {Lanza}, {Pagano}, {Fares}, {Shkolnik}, {Bohlender}, {Cameron}, {Dieters}, {Donati}, {Mart{\'{\i}}nez Fiorenzano}, {Jardine}, and {Moutou}}]{scandariato2013}
{Scandariato} G, {Maggio} A, {Lanza} AF et~al. (2013) {A coordinated optical and X-ray spectroscopic campaign on HD 179949: searching for planet-induced chromospheric and coronal activity}. \aap 552:A7

\bibitem[{{Schr{\"o}ter} et~al.(2011){Schr{\"o}ter}, {Czesla}, {Wolter}, {M{\"u}ller}, {Huber}, and {Schmitt}}]{schro2011}
{Schr{\"o}ter} S, {Czesla} S, {Wolter} U et~al. (2011) {The corona and companion of CoRoT-2a. Insights from X-rays and optical spectroscopy}. \aap 532:A3

\bibitem[{{Shkolnik} et~al.(2003){Shkolnik}, {Walker}, and {Bohlender}}]{shkolnik2003}
{Shkolnik} E, {Walker} GAH {Bohlender} DA (2003) {Evidence for Planet-induced Chromospheric Activity on HD 179949}. \apj 597:1092--1096

\bibitem[{{Shkolnik} et~al.(2005{\natexlab{a}}){Shkolnik}, {Walker}, {Bohlender}, {Gu}, and {K{\"u}rster}}]{shkolnik2005}
{Shkolnik} E, {Walker} GAH, {Bohlender} DA, {Gu} PG {K{\"u}rster} M (2005{\natexlab{a}}) {Hot Jupiters and Hot Spots: The Short- and Long-Term Chromospheric Activity on Stars with Giant Planets}. \apj 622:1075--1090

\bibitem[{{Shkolnik} et~al.(2005{\natexlab{b}}){Shkolnik}, {Walker}, {Rucinski}, {Bohlender}, and {Davidge}}]{shkolnik2005b}
{Shkolnik} E, {Walker} GAH, {Rucinski} SM, {Bohlender} DA {Davidge} TJ (2005{\natexlab{b}}) {Investigating Ca II Emission in the RS Canum Venaticorum Binary ER Vulpeculae Using the Broadening Function Formalism}. \aj 130:799--808

\bibitem[{{Shkolnik} et~al.(2008){Shkolnik}, {Bohlender}, {Walker}, and {Collier Cameron}}]{shkolnik2008}
{Shkolnik} E, {Bohlender} DA, {Walker} GAH {Collier Cameron} A (2008) {The On/Off Nature of Star-Planet Interactions}. \apj 676:628-638

\bibitem[{{Shkolnik}(2013)}]{shko13}
{Shkolnik} EL (2013) {An Ultraviolet Investigation of Activity on Exoplanet Host Stars}. \apj 766:9

\bibitem[{{Strugarek}(2016)}]{strugarek2016}
{Strugarek} A (2016) {Assessing Magnetic Torques and Energy Fluxes in Close-in Star-Planet Systems}. \apj 833(2):140

\bibitem[{{Strugarek} et~al.(2022){Strugarek}, {Fares}, {Bourrier}, {Brun}, {R{\'e}ville}, {Amari}, {Helling}, {Jardine}, {Llama}, {Moutou}, {Vidotto}, {Wheatley}, and {Zarka}}]{strugarek2023}
{Strugarek} A, {Fares} R, {Bourrier} V et~al. (2022) {MOVES - V. Modelling star-planet magnetic interactions of HD 189733}. \mnras 512(3):4556--4572

\bibitem[{{T{\'o}th} et~al.(2005){T{\'o}th}, Sokolov, Gombosi, Chesney, Clauer, De~Zeeuw, Hansen, Kane, Manchester, Oehmke, Powell, Ridley, Roussev, Stout, Volberg, Wolf, Sazykin, Chan, Yu, and Kóta}]{toth2005}
{T{\'o}th} G, Sokolov IV, Gombosi TI et~al. (2005) Space weather modeling framework: A new tool for the space science community. Journal of Geophysical Research: Space Physics 110(A12):n/a--n/a, \urlprefix\url{http://dx.doi.org/10.1029/2005JA011126}, a12226

\bibitem[{{T{\'o}th} et~al.(2012){T{\'o}th}, {van der Holst}, {Sokolov}, {De Zeeuw}, {Gombosi}, {Fang}, {Manchester}, {Meng}, {Najib}, {Powell}, {Stout}, {Glocer}, {Ma}, and {Opher}}]{toth2012}
{T{\'o}th} G, {van der Holst} B, {Sokolov} IV et~al. (2012) {Adaptive numerical algorithms in space weather modeling}. Journal of Computational Physics 231:870--903

\bibitem[{{Treumann}(2006)}]{treumann2006}
{Treumann} RA (2006) {The electron-cyclotron maser for astrophysical application}. \aapr 13:229--315

\bibitem[{{Trigilio} et~al.(2023){Trigilio}, {Biswas}, {Leto}, {Umana}, {Busa}, {Cavallaro}, {Das}, {Chandra}, {Perez-Torres}, {Wade}, {Bordiu}, {Buemi}, {Bufano}, {Ingallinera}, {Loru}, and {Riggi}}]{trigilio2023}
{Trigilio} C, {Biswas} A, {Leto} P et~al. (2023) {Star-Planet Interaction at radio wavelengths in YZ Ceti: Inferring planetary magnetic field}. arXiv e-prints arXiv:2305.00809

\bibitem[{Turner et~al.(2021)Turner, Zarka, Grie{\ss}meier, Lazio, Cecconi, Enriquez, Girard, Jayawardhana, Lamy, Nichols, and de~Pater}]{turner2021}
Turner JD, Zarka P, Grie{\ss}meier JM et~al. (2021) The search for radio emission from the exoplanetary systems 55 {{Cancri}}, {$\upsilon$} {{Andromedae}}, and {$\tau$} {{Bo\"otis}} using {{LOFAR}} beam-formed observations. Astronomy \& Astrophysics 645:A59

\bibitem[{{Vidotto} et~al.(2009){Vidotto}, {Opher}, {Jatenco-Pereira}, and {Gombosi}}]{vidotto2009}
{Vidotto} AA, {Opher} M, {Jatenco-Pereira} V {Gombosi} TI (2009) {Three-dimensional Numerical Simulations of Magnetized Winds of Solar-like Stars}. \apj 699:441--452

\bibitem[{{Vidotto} et~al.(2010){Vidotto}, {Jardine}, and {Helling}}]{vidotto2010}
{Vidotto} AA, {Jardine} M {Helling} C (2010) {Early UV Ingress in WASP-12b: Measuring Planetary Magnetic Fields}. \apjl 722:L168--L172

\bibitem[{{Vidotto} et~al.(2012){Vidotto}, {Fares}, {Jardine}, {Donati}, {Opher}, {Moutou}, {Catala}, and {Gombosi}}]{vidotto2012}
{Vidotto} AA, {Fares} R, {Jardine} M et~al. (2012) {The stellar wind cycles and planetary radio emission of the {$\tau$} Boo system}. \mnras 423:3285--3298

\bibitem[{{Vidotto} et~al.(2013){Vidotto}, {Jardine}, {Morin}, {Donati}, {Lang}, and {Russell}}]{vidotto2013}
{Vidotto} AA, {Jardine} M, {Morin} J et~al. (2013) {Effects of M dwarf magnetic fields on potentially habitable planets}. \aap 557:A67

\bibitem[{{Vidotto} et~al.(2014){Vidotto}, {Jardine}, {Morin}, {Donati}, {Opher}, and {Gombosi}}]{vidotto2014}
{Vidotto} AA, {Jardine} M, {Morin} J et~al. (2014) {M-dwarf stellar winds: the effects of realistic magnetic geometry on rotational evolution and planets}. \mnras 438:1162--1175

\bibitem[{{Vidotto} et~al.(2023){Vidotto}, {Bourrier}, {Fares}, {Bellotti}, {Donati}, {Petit}, {Hussain}, and {Morin}}]{vidotto2023}
{Vidotto} AA, {Bourrier} V, {Fares} R et~al. (2023) {The space weather around the exoplanet GJ 436b. II. Stellar wind-exoplanet interactions}. \aap 678:A152

\bibitem[{{Walker} et~al.(2008){Walker}, {Croll}, {Matthews}, {Kuschnig}, {Huber}, {Weiss}, {Shkolnik}, {Rucinski}, {Guenther}, {Moffat}, and {Sasselov}}]{walker2008}
{Walker} GAH, {Croll} B, {Matthews} JM et~al. (2008) {MOST detects variability on {$\tau$} Bootis A possibly induced by its planetary companion}. \aap 482:691--697

\bibitem[{{Wood} et~al.(2005){Wood}, {M{\"u}ller}, {Zank}, {Linsky}, and {Redfield}}]{wood2005}
{Wood} BE, {M{\"u}ller} HR, {Zank} GP, {Linsky} JL {Redfield} S (2005) {New Mass-Loss Measurements from Astrospheric Ly{$\alpha$} Absorption}. \apjl 628:L143--L146

\bibitem[{Yadav and Thorngren(2017)}]{yadav2017}
Yadav RK Thorngren DP (2017) Estimating the {{Magnetic Field Strength}} in {{Hot Jupiters}}. The Astrophysical Journal 849(1):L12

\bibitem[{{Zarka} et~al.(2001){Zarka}, {Treumann}, {Ryabov}, and {Ryabov}}]{zarka2001}
{Zarka} P, {Treumann} RA, {Ryabov} BP {Ryabov} VB (2001) {Magnetically-Driven Planetary Radio Emissions and Application to Extrasolar Planets}. \apss 277:293--300

\end{thebibliography}

\end{document}